\documentclass[revtex,prb,reprint,superscriptaddress]{revtex4-1}

\usepackage{amsmath}
\usepackage{amsfonts}
\usepackage{amssymb}
\usepackage{graphicx}
\usepackage{moreverb}
\usepackage{url}
\usepackage{epsf}
\usepackage{color}
\usepackage{multirow}
\usepackage{subfigure}
\usepackage{hyperref}
\usepackage{soul}
\DeclareMathOperator{\sgn}{sgn}

\begin{document}

\title{First-principles multiscale modelling of charged adsorbates on doped graphene}
\author{Fabiano~Corsetti}
\affiliation{Departments of Materials and Physics, and the Thomas Young Centre for Theory and Simulation of Materials, Imperial College London, London SW7 2AZ, United Kingdom}
\author{Arash~A.~Mostofi}
\affiliation{Departments of Materials and Physics, and the Thomas Young Centre for Theory and Simulation of Materials, Imperial College London, London SW7 2AZ, United Kingdom}
\author{Johannes~Lischner}
\email[E-mail: ]{j.lischner@imperial.ac.uk}
\affiliation{Departments of Materials and Physics, and the Thomas Young Centre for Theory and Simulation of Materials, Imperial College London, London SW7 2AZ, United Kingdom}
\date{\today}

\begin{abstract}
Adsorbed atoms and molecules play an important role in controlling and tuning the functional properties of two-dimensional (2D) materials. Understanding and predicting this process from theory is challenging because of the need to capture the complex interplay between the local chemistry and the long-range screening response. To address this problem, we present a first-principles multiscale approach that combines linear-scaling density-functional theory, continuum screening theory and large-scale tight-binding simulations into a seamless parameter-free theory of adsorbates on 2D materials. We apply this method to investigate the electronic structure of doped graphene with a single calcium (Ca) adatom and find that the Ca atom acts as a Coulomb impurity which modifies the graphene local density of states (LDOS) within a distance of several nanometres in its vicinity. We also observe an important doping dependence of the graphene LDOS near the Ca atom, which gives insights into electronic screening in graphene. Our multiscale framework opens up the possibility of investigating complex mesoscale adsorbate configurations on 2D materials relevant to real devices.
\end{abstract}

\maketitle

\section{Introduction}

Adsorbate engineering is a promising route towards modifying the electronic structure of graphene and other 2D materials for emerging functional device technologies~\cite{graphene_book,Schedin2007,Szafranek2011,Ludbrook22092015}. For example, adsorbed atoms and molecules can change the carrier mobility~\cite{Adam2009,DasSarma2011}, effective dielectric constant~\cite{Jang2008}, chemical potential~\cite{Zhao999} and local magnetic moment~\cite{Gonzalez-Herrero437} of graphene.

Charge transfer from adsorbates to the graphene substrate results in doping and the creation of Coulomb impurities. From a theoretical perspective, the description of such impurities and their interaction with the graphene substrate constitutes a challenging multiscale problem: while the charge $Z$ transferred between the adsorbate and substrate is principally determined by the local chemistry (i.e., the coupling between adsorbate and substrate orbitals), the screening response of graphene is long-ranged with an extent of several nanometres~\cite{Wang2012}. Experimentally, the structural properties and screening response of adsorbed species on graphene as well as other point defects can be investigated with scanning probe techniques, such as scanning tunneling microscopy and spectroscopy or atomic force microscopy~\cite{Zhao999,Brar2011,Wang2012,Wang2013,PhysRevLett.112.036804,Mao2016}.

Previous theoretical studies of adsorbates on graphene have focussed either on the description of the short-range charge transfer or the long-range graphene response. First-principles density-functional theory (DFT) has been used to determine the charge transfer for different adatom species; however, the supercells in these calculations were much too small to capture the long-range screening response of graphene~\cite{Chan2008,Parq2010,Chan2011,Sun2015668}. Conversely, theoretical calculations based on tight-binding (TB) or continuum Dirac models~\cite{Pereira2007,Wang2012,Polini2008,Lu2015}, while able to access large length scales, require external parameters (most importantly, the impurity charge $Z$) and so do not possess the unbiased predictive power of first-principles methods. Furthermore, such models typically employ uncontrolled approximations for the description of electron-electron interactions, which are either completely neglected~\cite{Pereira2007,Novikov2007,Wang2012,Wang2013} or treated within linear-response~\cite{Ando2006,Hwang2007} or Thomas-Fermi (TF) theory~\cite{Katsnelson2006,Fogler2007,Terekhov2008}. Thus, there is a need for a parameter-free theory capable of bridging the length scales relevant to graphene-adsorbate interactions.

In this study, we combine first-principles DFT, continuum screening theory and model TB Hamiltonians within a novel parameter-free multiscale approach in order to overcome the limitations of each method in isolation. We demonstrate the power of our approach by applying it to the case of an isolated Ca atom adsorbed on doped graphene. This is a case of particular interest due to ongoing work on the detailed experimental characterization of the doping-dependent screening response~\cite{WongCa}, and the observation of supercritical states for small clusters of Ca adatom dimers~\cite{Wang2013}.

Ca donates electrons to the graphene resulting in an impurity charge of $Z = 1.6$~e (with e denoting the proton charge). We compute the graphene local density of states (LDOS) in the vicinity of the charged Ca atom and observe that it decays to the bulk graphene value on a length scale of several nanometres. The decay length depends on the chemical potential which influences graphene's screening response to the impurity potential.

The remainder of this article is structured as follows: in Sec.~\ref{sec:methods} we describe our general multiscale approach that bridges large-scale first-principles electronic structure calculations, continuum models and TB simulations at the mesoscale, and details of how we apply this approach to the specific case of a Ca adsorbate on a graphene substrate; in Sec.~\ref{sec:results} we present and discuss our results; and we make some concluding remarks in Sec.~\ref{sec:conclusions}.

\section{Methods\label{sec:methods}}

\subsection{Outline of multiscale approach\label{sec:outline}}

\begin{figure}[t]
\includegraphics[width=0.44\textwidth]{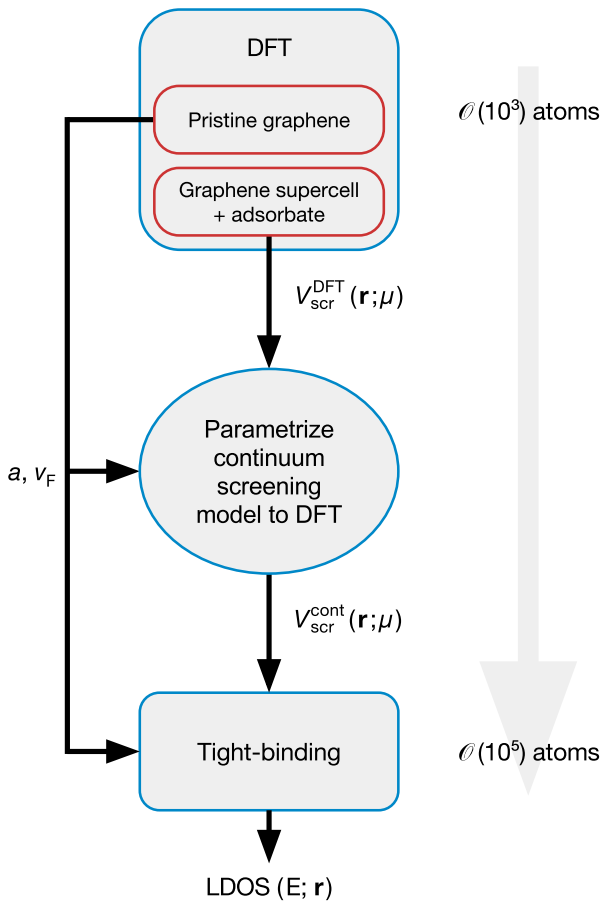}
\caption{Schematic diagram of the multiscale method used to simulate the long-range screening of a charged adsorbate on graphene. The intrinsic properties of graphene (the lattice constant $a$ and Fermi velocity $v_\mathrm{F}$) are extracted from first-principles DFT simulations of the pristine crystal. The screened potential $V_\mathrm{scr} \left ( \mathbf{r}; \mu \right)$ due to the adsorbate is calculated using a continuum screening model; the model depends on the charge transfer $Z$, which is obtained by fitting the model potential to the first-principles self-consistent potential extracted from DFT simulations of the graphene with adsorbate system.}
\label{fig:flowchart}
\end{figure}

Fig.~\ref{fig:flowchart} shows the multiscale approach we developed to model the properties of adsorbates on doped graphene. The method consists of three key steps. First, we perform first-principles DFT calculations of the pristine graphene and the graphene with adsorbed species. From the former, we derive the pristine graphene lattice parameter $a$ and the Fermi velocity $v_\mathrm{F}$, which we use in the continuum and TB Hamiltonians of the second and third steps. The periodic supercell for the graphene with adsorbate must be sufficiently large to converge chemical interactions between the adsorbate and the substrate~\cite{Bell2014}. To sample different values of the carrier density and, thereby, chemical potential $\mu$, we perform a series of calculations with different numbers of electrons in the supercell.

Secondly, we parametrize a continuum screening model of the graphene with the adsorbed species. In this approach, the adsorbate is treated as a point impurity charge $Z$ located above the graphene sheet, which is described by the Dirac Hamiltonian (i.e. without atomistic detail). To determine the charge transfer between the adsorbate and graphene, we compute the self-consistent (or screened) potential of the continuum model in the same supercell that was used for the DFT calculations and, for each value of $\mu$, we fit the value of $Z$ to achieve optimal agreement with the DFT self-consistent potential. Different continuum screening models may be used in this procedure, corresponding to different treatments of electron-electron interactions. In this work we employ non-linear Thomas-Fermi theory~\cite{Katsnelson2006} (NLTF), but also include interband transitions through an effective dielectric function. The screened potential at a position $\mathbf{r}$ in the graphene plane resulting from an impurity at $\mathbf{R} = \left ( 0, 0, z_\mathrm{imp} \right )$ is given by 
\begin{equation} \label{eq:Vcont}
V^\mathrm{cont}_\mathrm{scr} \left ( \mathbf{r}; \mu \right ) = Z v_{z_\mathrm{imp}}(\mathbf{r}) + \int d^2 \mathbf{r}' [n(\mathbf{r}')-n_0]v_0(\mathbf{r}-\mathbf{r}'),
\end{equation}
where $n_0 \left ( \mu \right ) = \mu \left | \mu \right | / \left ( \pi v^2_\mathrm{F} \right )$ and $n \left ( \mathbf{r} \right ) = n_0 \left ( \mu - V_\mathrm{scr} \left ( \mathbf{r} \right ) \right )$ denote the electron density of graphene without and with the adsorbate, respectively, and $v_z \left ( \mathbf{r} \right )$ is the Coulomb interaction screened by graphene interband transitions, i.e. $v_z \left ( q \right ) = \epsilon^{-1}_\mathrm{inter} \left ( q \right ) 2 \pi / q \times \exp{\left ( -qz \right )}$ with $\epsilon^{-1}_\mathrm{inter} \left ( q \right )$ being the interband contribution to the graphene dielectric function in the random-phase approximation~\cite{Shung1986,Wunsch2006,Hwang2007b}. In the above equation, the first term captures the bare impurity potential and the second term is the induced potential arising from electronic screening. We denote this model, which reduces to the well-known random-phase approximation in the linear-response limit~\cite{Shung1986,Wunsch2006,Hwang2007b}, as NLTF+inter. We further included additional screening due to graphene $\sigma$ bands~\cite{Wehling2011} and exchange-correlation effects~\cite{Polini2008}, but found that they have a small effect and can be neglected.

Thirdly, we perform a TB calculation in a supercell that is sufficiently large to capture the decay of the screened impurity potential. We find that the supercells needed to converge the screened potential are several orders of magnitude larger than the supercells needed to converge chemical interactions. The screened potential calculated from the continuum model, with the value of $Z$ that has been fitted to DFT, is used as an on-site term in the TB Hamiltonian. The TB calculation then yields the local density of states (LDOS), which can be compared to scanning tunneling experiments~\cite{Wang2012,WongCa}.

\subsection{Application to calcium adatoms on graphene}

In this section, we describe the application of the multiscale approach to isolated Ca adatoms on doped graphene. 

{\bf Structure ---} All calculations use hexagonal supercells containing $n \times n$ graphene unit cells ($2 n^2$ C atoms) and a single Ca adatom. We use a graphene lattice constant $a$ of 2.47~\AA, obtained from plane-wave DFT calculations of a single unit cell with an energy cutoff of 1000~eV, a $28 \times 28$ k-point grid (centred at the $\Gamma$-point), norm-conserving pseudopotentials, 20~\AA~separation between periodic images of the graphene sheets and the PBE exchange-correlation energy functional~\cite{PBE}. These calculations were carried out using the CASTEP~\cite{clark2005first} code. For the Ca adatom we use the geometry found by a previous DFT study of the system~\cite{Chan2008}, with the adatom 2~\AA\ above the hollow site at the centre of a hexagon. The same study also reported negligibly small relaxations of the graphene atoms near the Ca adatom and we thus keep these atoms fixed in our calculations.  

{\bf First-principles calculations ---} We perform DFT calculations in a $56 \times 56$ graphene supercell (6,272 C atoms and a single Ca adatom). While this supercell is not large enough to capture the decay of the screened impurity potential, it is sufficiently large to describe chemical interactions resulting from the coupling of adsorbate and graphene states, such as exchange interactions giving rise to spin-split states. Such calculations are already very large by the standards of DFT, and are made possible by using the {\sc ONETEP}~\cite{ONETEP,LDOS} DFT code whose computational cost scales only linearly with the size of the system, in contrast to the cubic scaling of conventional methods. We use norm-conserving pseudopotentials (including the 3s and 3p semi-core states for the Ca atom) and the PBE exchange-correlation functional. The Brillouin zone is sampled at the $\Gamma$-point. Kohn-Sham states are expanded in a basis of non-orthogonal, atom-centred orbitals with a radius of 5.3~\AA\ that are variationally optimized {\em in situ}, and described on a real-space grid corresponding to a plane-wave energy cutoff of 1000~eV. As we are dealing with a semi-metallic system, we do not truncate the density matrix and employ an ensemble-DFT~\cite{EDFT} formalism with an electronic temperature of 50~K. The simulation employs periodic boundary conditions in all three directions, and we use a periodic image separation perpendicular to the graphene plane of 20~\AA.

In experiments on graphene in a field-effect transistor setup the density of carriers can be modified by application of a gate voltage. To explore the effect of varying the carrier density in our simulations, we perform separate DFT calculations with different numbers of electrons in the supercell. Specifically, by adding and subtracting pairs of electrons we vary the total system charge $\Delta Q$ from $-4$ e to $+8$ e, corresponding to carrier densities $n$ between $-3 \times 10^{12}$~cm$^{-2}$ and $5 \times 10^{12}$~cm$^{-2}$. For calculations with a non-zero total charge $\Delta Q$ we use a compensating homogeneous background charge. 

The self-consistent (screened) local potential from each DFT calculation, which is comprised of the ionic, Hartree and exchange-correlation terms, is locally averaged in a Voronoi cell around each C atom in the graphene plane in order to smooth out the large variations that occur on the scale of an interatomic spacing. This smoothed DFT potential $V^\mathrm{DFT}_\mathrm{scr} \left ( \mathbf{r}; \Delta Q \right)$ is then used to parametrize the continuum model.

{\bf Continuum screening calculations ---} The screened impurity potential from the continuum screening model is calculated with a custom in-house code with 2D periodicity~\cite{graphene-code}. The centre of the impurity potential is placed 2~\AA\ above a hollow site at the centre of a hexagon of the graphene plane, as in the DFT geometry. For the non-linear models, a self-consistency cycle is used with a mixing scheme in which the screened potential of the next iteration is composed of 1\% of the screened potential of the current iteration and of 99\% of the screened potential from the previous iteration. 

The continuum screening model depends on three parameters: $Z$, $\mu$ and the Fermi velocity $v_\mathrm{F}$. The substrate dielectric constant usually constitutes an additional parameter; here, however, we are interested in suspended graphene and so set it to unity. $v_\mathrm{F} = 0.8 \times 10^6$~m/s is taken directly from the pristine graphene DFT band structure, and $Z$ and $\mu$ are treated instead as fitting parameters of the model. 

The accuracy of the model is evaluated against DFT using a fitness metric $\mathcal{F}(Z,\mu)$, which is defined as the root-mean-square difference between the model potential $V^\mathrm{cont}_\mathrm{scr}$ of Eq.~\eqref{eq:Vcont} and the DFT potential $V^\mathrm{DFT}_\mathrm{scr}$, integrated over the supercell area. We calculate $V^\mathrm{cont}_\mathrm{scr}$ for a range of $Z$ and $\mu$ values using the same supercell as in the DFT calculation (including periodic boundary conditions). The values of $Z$ and $\mu$ that minimize $\mathcal{F}$ are those that give the optimal fit of the screening model to DFT (see Fig.~S1 of the Supplementary Material).

To study the effect of different approximations for the electron-electron interactions, we also use three other continuum screening models: (i) TF theory within a linear-response formalism (LTF), which only includes intraband transitions between the Dirac bands of graphene; (ii) linear-response TF theory plus interband transitions (LTF+inter); (iii) non-linear TF theory (NLTF)~\cite{Katsnelson2006} without interband transitions.

{\bf Tight-binding calculations ---} The large-scale TB simulations are performed in a $168 \times 168$ supercell (56,448 C atoms and a single Ca adatom); this is an order of magnitude increase in system size with respect to the underlying DFT simulations. We use a custom in-house TB code for graphene with 2D periodicity. The TB Hamiltonian includes the C p$_z$ orbitals and nearest-neighbour hoppings only with a hopping energy of 2.54~eV. A $2 \times 2$ Monkhorst-Pack mesh (centred at the $\Gamma$-point) is used to sample the Brillouin zone.

\section{Results\label{sec:results}}

\subsection{Charge transfer}

\begin{figure}[t]
\includegraphics[width=0.49\textwidth]{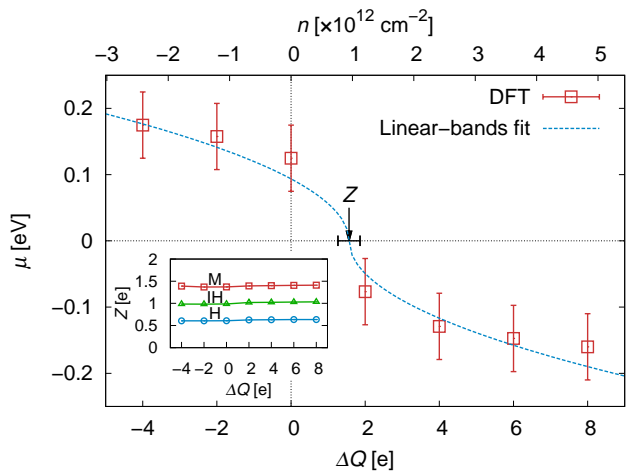}
\caption{DFT simulations of a Ca adatom in a $56 \times 56$ supercell of graphene. The main panel shows the chemical potential $\mu$ relative to the Dirac point as a function of supercell charge $\Delta Q$, and the fit to a linear-bands model of doped graphene. The error bars are determined by the spacing of DFT eigenstates in the energy spectrum, resulting from the finite supercell size and Brillouin zone sampling. The inset shows the charge transfer $Z$ obtained from three charge-partitioning schemes: Mulliken (M), Hirshfeld (H), Iterative Hirshfeld (IH). For the H and IH schemes we use a two-component system (the Ca atom and the graphene substrate) from which the pro-molecular density is constructed.}
\label{fig:mu_fit}
\end{figure}

The charge transfer $Z$ is an important quantity for characterizing the interaction between an adsorbate and the substrate. Standard approaches for calculating $Z$ partition the continuous electron density obtained from DFT into an adsorbate and a substrate contribution. For example, Chan {\em et al.}~\cite{Chan2008} calculated the charge transfer for different adatoms on graphene, but found large discrepancies between two different methods of partitioning the electron density (0.18~e and 0.78~e for Ca adatoms). As shown in the inset of Fig.~\ref{fig:mu_fit} and Fig.~S2 in the Supplementary Material, we obtain an even broader range of values for $Z$, from 0.2~e to 1.4~e, depending on the particular charge partitioning method used. While each method provides a different value for $Z$, all of them find the charge transfer to be independent of the supercell charge $\Delta Q$, indicating that Ca states are neither filled nor depleted as the graphene doping level is varied.

In contrast to these approaches, no partitioning of the charge density is required in our multiscale method. Instead, $Z$ is obtained by imposing that the screened impurity potential of the continuum screening model reproduces the self-consistent DFT potential. In the fitting procedure, we enforce a $\mu$-independent value of $Z$ by performing a global fit to all doping levels at once with equal weighting. This is justified by the charge stability of the impurity discussed previously. It should be noted that performing individual fits for each doping level indeed results in a fairly constant $Z$ for all charge states except $\Delta Q = 2$~e, in which case the value is enhanced by up to $\sim$50\%; we believe this to be because the graphene system is very close to being undoped and, hence, is the least well-described by our models as the screening is dominated in this case by the interband transitions~\cite{DiVincenzo1984}.

Table~\ref{table:models_fit} gives the value of $Z$ obtained from the new multiscale approach using different approximations for the description of electron-electron interactions. The NLTF+inter method, which offers the most accurate description of electron-electron interactions, yields $Z = 1.6$~e. The same value is obtained from the NLTF method indicating that inclusion of interband transition does not change the value of $Z$. In contrast, the linear-response LTF and LTF+inter approaches yield significantly smaller values for $Z$. 

Table~\ref{table:models_fit} also gives the value of the fitness function $\mathcal{F}$ averaged over all DFT doping levels. It is interesting to note that there is almost no difference in this value between the different approaches~\footnote{However, the four models result in somewhat lower overall values of $\mathcal{F}$ than fits based on empirical forms for the screened potential, discussed in Fig.~S3 in the Supplementary Material.}; all of them reproduce the DFT screened potential with high accuracy ($< 10$~meV), but with significantly different $Z$ values. This observation was explained by Katsnelson~\cite{Katsnelson2006} who pointed out that the screened potential of a charged impurity in doped graphene has the same functional form in linear and non-linear Thomas-Fermi theory, but with a different prefactor corresponding to different effective values of $Z$.

An alternative method for calculating $Z$ which also does not require a partitioning of the charge density is shown in the main panel of Fig.~\ref{fig:mu_fit}. For each value of the supercell charge $\Delta Q$, we determine $\mu$ from the spectrum of Kohn-Sham energies (open squares) and then fit a linear-bands model of doped graphene of the form\break $\mu \left ( \Delta Q \right ) = -\sgn{\left [ \Delta Q - Z \right ]} v_\mathrm{F} \sqrt{\left | \Delta Q - Z \right | \pi / A}$ (dashed line), where $\Delta Q - Z$ is the total free carrier charge and $A$ is the area of the supercell. This method yields $Z = 1.6 \pm 0.3$~e, significantly higher than previously suggested values~\cite{Chan2008}, but in excellent agreement with our multiscale method. 

\begin{table}
\caption{Charge transfer $Z$ obtained from the new multiscale approach, i.e. by fitting the screened impurity potential of a continuum screening model to the self-consistent DFT potential, and the corresponding value of the fitness metric $\mathcal{F}$ of the model averaged over all DFT doping levels. We also include the value of the adatom charge estimated from the DFT DOS (see Fig.~\ref{fig:mu_fit}).}
\label{table:models_fit}
\begin{ruledtabular}
{\footnotesize
\begin{tabular*}{\textwidth}{lcc}
Model                                           & $Z$ [e]       & $\mathcal{F}$ [eV] \\
\hline
LTF                                             & 0.3           & $8.9 \times 10^{-3}$ \\
LTF+inter                                       & 1.3           & $8.4 \times 10^{-3}$ \\
NLTF                                            & 1.6           & $9.3 \times 10^{-3}$ \\
NLTF+inter                                      & 1.6           & $7.4 \times 10^{-3}$ \\
\hline
Linear-bands fit to DFT (Fig.~\ref{fig:mu_fit}) & $1.6 \pm 0.3$ & -
\end{tabular*}
}
\end{ruledtabular}
\end{table}

\subsection{Local density of states} \label{subsec:STM}

\begin{figure}[t]
\includegraphics[width=0.49\textwidth]{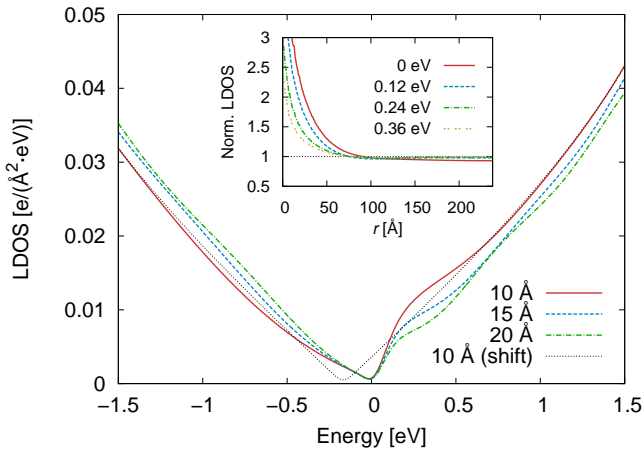}
\caption{LDOS of free-standing, n-doped graphene at different distances from a single Ca adatom, calculated with the first-principles multiscale approach described in the text. The dotted line shows the LDOS of unperturbed graphene shifted in energy by the average value of the on-site screened potential at a distance of 10~\AA\ from the adatom. The inset shows the LDOS at 0.1~eV above the Dirac point as a function of the distance $r$ from the adatom (normalized by the value for unperturbed graphene), and as the graphene doping is increased (the chemical potential $\mu$ is given in the key).}
\label{fig:large_TB}
\end{figure}

Fig.~\ref{fig:large_TB} shows the LDOS obtained from the large-scale TB simulation with the screened potential from the NLTF+inter model. The main panel shows a representative example for the case of n-doped graphene (at a chemical potential $\mu = 0.12$~eV).

As observed in previous studies employing an unscreened impurity potential~\cite{Pereira2007,Wang2012}, the LDOS near the adatom exhibits a broken electron-hole symmetry. This behavior can be understood within a simple TF-based picture, in which the LDOS in the presence of the adsorbate is given by that of unperturbed graphene, but shifted in energy by the value of the screened impurity potential (shown by the dotted line in Fig.~\ref{fig:large_TB}). While the shifted LDOS agrees well with the full result for energies more than $\pm \sim$0.5~eV from the Dirac point, there are significant deviations within this range. In particular, the position of the Dirac point remains pinned at its unperturbed value, which is a consequence of the linear dispersion of the bands from graphene's chiral Dirac fermions.

The inset of Fig.~\ref{fig:large_TB} shows the decay of the LDOS at a fixed energy towards its unperturbed value far from the adatom (corresponding to experimental d$I$/d$V$ linescans~\cite{Wang2012}). The decay length decreases with increasing $\left | \mu \right |$. This is in agreement with very recent experimental measurements~\cite{WongCa}. The chemical potential only enters the simulation through its effect on the screened potential; this demonstrates the impact of screening on measured properties of the system and highlights the importance of an accurate description of electron-electron interactions.

\section{Conclusions\label{sec:conclusions}}

In conclusion, we have introduced an accurate and efficient multiscale approach for the theoretical description of charged impurities on doped graphene. Starting from linear-scaling DFT simulations of a Ca adatom in a $56 \times 56$ graphene supercell, we parametrize a continuum screening model by fitting the screened potential of the model to the DFT self-consistent potential. The continuum screening model is then evaluated in a much larger $168 \times 168$ supercell and the resulting screened impurity potential is used as input for large-scale TB simulations, which yield observable quantities, such as the local density of states.

Although the continuum screening model is non-analytic and requires a self-consistent calculation, its computational cost is negligible, many orders of magnitude less than the TB simulation. The TB simulation itself is also computationally inexpensive: the $168 \times 168$ supercell with TB is $\sim$0.1\% of the cost of the $56 \times 56$ supercell with DFT, allowing us to access much larger systems than would be possible with DFT alone. As shown in Fig.~S4 in the Supplementary Material, the limiting factor in the accuracy of the large-scale simulation is the TB approximation itself rather than our fitting procedure.

This multiscale approach, which bridges local chemistry at the adsorbate site on the nanoscale and long-range screening effects on the mesoscale within a single, physically-motived framework, opens up the possibility of investigating more realistic and complex mescoscale configurations relevant for the understanding of real devices, e.g., disordered arrangements of different adsorbate species. 

Data underlying the DFT calculations are available on figshare~\cite{figshare-files} and may be used under the Creative Commons Attribution licence. The code for calculating the screened potential of such systems with the NLTF+inter model is available online with a user-friendly web-based interface~\cite{graphene-code}.

\begin{acknowledgments}
This work was supported by the Thomas Young Centre under grant TYC-101. The UK Engineering and Physical Sciences Research Council (EPSRC) supported FC and AAM under grant EP/J015059/1, and JL under grant EP/N005244/1. The calculations were performed on cx1/Helen (Imperial College High Performance Computing Service) and ARCHER (UK National Supercomputing Service). We thank Dillon Wong for helpful discussions, and the UK's HEC Materials Chemistry Consortium for access to ARCHER under grant EP/L000202.
\end{acknowledgments}


\begin{thebibliography}{42}%
\makeatletter
\providecommand \@ifxundefined [1]{%
 \@ifx{#1\undefined}
}%
\providecommand \@ifnum [1]{%
 \ifnum #1\expandafter \@firstoftwo
 \else \expandafter \@secondoftwo
 \fi
}%
\providecommand \@ifx [1]{%
 \ifx #1\expandafter \@firstoftwo
 \else \expandafter \@secondoftwo
 \fi
}%
\providecommand \natexlab [1]{#1}%
\providecommand \enquote  [1]{``#1''}%
\providecommand \bibnamefont  [1]{#1}%
\providecommand \bibfnamefont [1]{#1}%
\providecommand \citenamefont [1]{#1}%
\providecommand \href@noop [0]{\@secondoftwo}%
\providecommand \href [0]{\begingroup \@sanitize@url \@href}%
\providecommand \@href[1]{\@@startlink{#1}\@@href}%
\providecommand \@@href[1]{\endgroup#1\@@endlink}%
\providecommand \@sanitize@url [0]{\catcode `\\12\catcode `\$12\catcode
  `\&12\catcode `\#12\catcode `\^12\catcode `\_12\catcode `\%12\relax}%
\providecommand \@@startlink[1]{}%
\providecommand \@@endlink[0]{}%
\providecommand \url  [0]{\begingroup\@sanitize@url \@url }%
\providecommand \@url [1]{\endgroup\@href {#1}{\urlprefix }}%
\providecommand \urlprefix  [0]{URL }%
\providecommand \Eprint [0]{\href }%
\providecommand \doibase [0]{http://dx.doi.org/}%
\providecommand \selectlanguage [0]{\@gobble}%
\providecommand \bibinfo  [0]{\@secondoftwo}%
\providecommand \bibfield  [0]{\@secondoftwo}%
\providecommand \translation [1]{[#1]}%
\providecommand \BibitemOpen [0]{}%
\providecommand \bibitemStop [0]{}%
\providecommand \bibitemNoStop [0]{.\EOS\space}%
\providecommand \EOS [0]{\spacefactor3000\relax}%
\providecommand \BibitemShut  [1]{\csname bibitem#1\endcsname}%
\let\auto@bib@innerbib\@empty
\bibitem [{\citenamefont {Katsnelson}(2012)}]{graphene_book}%
  \BibitemOpen
  \bibfield  {author} {\bibinfo {author} {\bibfnamefont {M.~I.}\ \bibnamefont
  {Katsnelson}},\ }\href@noop {} {\emph {\bibinfo {title} {Graphene: Carbon in
  Two Dimensions}}}\ (\bibinfo  {publisher} {Cambridge University Press},\
  \bibinfo {address} {Cambridge},\ \bibinfo {year} {2012})\BibitemShut
  {NoStop}%
\bibitem [{\citenamefont {Schedin}\ \emph {et~al.}(2007)\citenamefont
  {Schedin}, \citenamefont {Geim}, \citenamefont {Morozov}, \citenamefont
  {Hill}, \citenamefont {Blake}, \citenamefont {Katsnelson},\ and\
  \citenamefont {Novoselov}}]{Schedin2007}%
  \BibitemOpen
  \bibfield  {author} {\bibinfo {author} {\bibfnamefont {F.}~\bibnamefont
  {Schedin}}, \bibinfo {author} {\bibfnamefont {A.~K.}\ \bibnamefont {Geim}},
  \bibinfo {author} {\bibfnamefont {S.~V.}\ \bibnamefont {Morozov}}, \bibinfo
  {author} {\bibfnamefont {E.~W.}\ \bibnamefont {Hill}}, \bibinfo {author}
  {\bibfnamefont {P.}~\bibnamefont {Blake}}, \bibinfo {author} {\bibfnamefont
  {M.~I.}\ \bibnamefont {Katsnelson}}, \ and\ \bibinfo {author} {\bibfnamefont
  {K.~S.}\ \bibnamefont {Novoselov}},\ }\href {\doibase 10.1038/nmat1967}
  {\bibfield  {journal} {\bibinfo  {journal} {Nature Mater.}\ }\textbf
  {\bibinfo {volume} {6}},\ \bibinfo {pages} {652} (\bibinfo {year}
  {2007})}\BibitemShut {NoStop}%
\bibitem [{\citenamefont {Szafranek}\ \emph {et~al.}(2011)\citenamefont
  {Szafranek}, \citenamefont {Schall}, \citenamefont {Otto}, \citenamefont
  {Neumaier},\ and\ \citenamefont {Kurz}}]{Szafranek2011}%
  \BibitemOpen
  \bibfield  {author} {\bibinfo {author} {\bibfnamefont {B.~N.}\ \bibnamefont
  {Szafranek}}, \bibinfo {author} {\bibfnamefont {D.}~\bibnamefont {Schall}},
  \bibinfo {author} {\bibfnamefont {M.}~\bibnamefont {Otto}}, \bibinfo {author}
  {\bibfnamefont {D.}~\bibnamefont {Neumaier}}, \ and\ \bibinfo {author}
  {\bibfnamefont {H.}~\bibnamefont {Kurz}},\ }\href {\doibase
  10.1021/nl200631m} {\bibfield  {journal} {\bibinfo  {journal} {Nano Lett.}\
  }\textbf {\bibinfo {volume} {11}},\ \bibinfo {pages} {2640} (\bibinfo {year}
  {2011})}\BibitemShut {NoStop}%
\bibitem [{\citenamefont {Ludbrook}\ \emph {et~al.}(2015)\citenamefont
  {Ludbrook}, \citenamefont {Levy}, \citenamefont {Nigge}, \citenamefont
  {Zonno}, \citenamefont {Schneider}, \citenamefont {Dvorak}, \citenamefont
  {Veenstra}, \citenamefont {Zhdanovich}, \citenamefont {Wong}, \citenamefont
  {Dosanjh}, \citenamefont {Straßer}, \citenamefont {Stöhr}, \citenamefont
  {Forti}, \citenamefont {Ast}, \citenamefont {Starke},\ and\ \citenamefont
  {Damascelli}}]{Ludbrook22092015}%
  \BibitemOpen
  \bibfield  {author} {\bibinfo {author} {\bibfnamefont {B.~M.}\ \bibnamefont
  {Ludbrook}}, \bibinfo {author} {\bibfnamefont {G.}~\bibnamefont {Levy}},
  \bibinfo {author} {\bibfnamefont {P.}~\bibnamefont {Nigge}}, \bibinfo
  {author} {\bibfnamefont {M.}~\bibnamefont {Zonno}}, \bibinfo {author}
  {\bibfnamefont {M.}~\bibnamefont {Schneider}}, \bibinfo {author}
  {\bibfnamefont {D.~J.}\ \bibnamefont {Dvorak}}, \bibinfo {author}
  {\bibfnamefont {C.~N.}\ \bibnamefont {Veenstra}}, \bibinfo {author}
  {\bibfnamefont {S.}~\bibnamefont {Zhdanovich}}, \bibinfo {author}
  {\bibfnamefont {D.}~\bibnamefont {Wong}}, \bibinfo {author} {\bibfnamefont
  {P.}~\bibnamefont {Dosanjh}}, \bibinfo {author} {\bibfnamefont
  {C.}~\bibnamefont {Straßer}}, \bibinfo {author} {\bibfnamefont
  {A.}~\bibnamefont {Stöhr}}, \bibinfo {author} {\bibfnamefont
  {S.}~\bibnamefont {Forti}}, \bibinfo {author} {\bibfnamefont {C.~R.}\
  \bibnamefont {Ast}}, \bibinfo {author} {\bibfnamefont {U.}~\bibnamefont
  {Starke}}, \ and\ \bibinfo {author} {\bibfnamefont {A.}~\bibnamefont
  {Damascelli}},\ }\href {\doibase 10.1073/pnas.1510435112} {\bibfield
  {journal} {\bibinfo  {journal} {Proc. Natl. Acad. Sci. U.S.A.}\ }\textbf
  {\bibinfo {volume} {112}},\ \bibinfo {pages} {11795} (\bibinfo {year}
  {2015})}\BibitemShut {NoStop}%
\bibitem [{\citenamefont {Adam}\ \emph {et~al.}(2009)\citenamefont {Adam},
  \citenamefont {Hwang}, \citenamefont {Rossi},\ and\ \citenamefont
  {Das~Sarma}}]{Adam2009}%
  \BibitemOpen
  \bibfield  {author} {\bibinfo {author} {\bibfnamefont {S.}~\bibnamefont
  {Adam}}, \bibinfo {author} {\bibfnamefont {E.~H.}\ \bibnamefont {Hwang}},
  \bibinfo {author} {\bibfnamefont {E.}~\bibnamefont {Rossi}}, \ and\ \bibinfo
  {author} {\bibfnamefont {S.}~\bibnamefont {Das~Sarma}},\ }\href {\doibase
  10.1016/j.ssc.2009.02.041} {\bibfield  {journal} {\bibinfo  {journal} {Solid
  State Commun.}\ }\textbf {\bibinfo {volume} {149}},\ \bibinfo {pages} {1072}
  (\bibinfo {year} {2009})}\BibitemShut {NoStop}%
\bibitem [{\citenamefont {Das~Sarma}\ \emph {et~al.}(2011)\citenamefont
  {Das~Sarma}, \citenamefont {Adam}, \citenamefont {Hwang},\ and\ \citenamefont
  {Rossi}}]{DasSarma2011}%
  \BibitemOpen
  \bibfield  {author} {\bibinfo {author} {\bibfnamefont {S.}~\bibnamefont
  {Das~Sarma}}, \bibinfo {author} {\bibfnamefont {S.}~\bibnamefont {Adam}},
  \bibinfo {author} {\bibfnamefont {E.~H.}\ \bibnamefont {Hwang}}, \ and\
  \bibinfo {author} {\bibfnamefont {E.}~\bibnamefont {Rossi}},\ }\href
  {\doibase 10.1103/RevModPhys.83.407} {\bibfield  {journal} {\bibinfo
  {journal} {Rev. Mod. Phys.}\ }\textbf {\bibinfo {volume} {83}},\ \bibinfo
  {pages} {407} (\bibinfo {year} {2011})}\BibitemShut {NoStop}%
\bibitem [{\citenamefont {Jang}\ \emph {et~al.}(2008)\citenamefont {Jang},
  \citenamefont {Adam}, \citenamefont {Chen}, \citenamefont {Williams},
  \citenamefont {Das~Sarma},\ and\ \citenamefont {Fuhrer}}]{Jang2008}%
  \BibitemOpen
  \bibfield  {author} {\bibinfo {author} {\bibfnamefont {C.}~\bibnamefont
  {Jang}}, \bibinfo {author} {\bibfnamefont {S.}~\bibnamefont {Adam}}, \bibinfo
  {author} {\bibfnamefont {J.-H.}\ \bibnamefont {Chen}}, \bibinfo {author}
  {\bibfnamefont {E.~D.}\ \bibnamefont {Williams}}, \bibinfo {author}
  {\bibfnamefont {S.}~\bibnamefont {Das~Sarma}}, \ and\ \bibinfo {author}
  {\bibfnamefont {M.~S.}\ \bibnamefont {Fuhrer}},\ }\href {\doibase
  10.1103/PhysRevLett.101.146805} {\bibfield  {journal} {\bibinfo  {journal}
  {Phys. Rev. Lett.}\ }\textbf {\bibinfo {volume} {101}},\ \bibinfo {pages}
  {146805} (\bibinfo {year} {2008})}\BibitemShut {NoStop}%
\bibitem [{\citenamefont {Zhao}\ \emph {et~al.}(2011)\citenamefont {Zhao},
  \citenamefont {He}, \citenamefont {Rim}, \citenamefont {Schiros},
  \citenamefont {Kim}, \citenamefont {Zhou}, \citenamefont {Guti{\'e}rrez},
  \citenamefont {Chockalingam}, \citenamefont {Arguello}, \citenamefont
  {P{\'a}lov{\'a}}, \citenamefont {Nordlund}, \citenamefont {Hybertsen},
  \citenamefont {Reichman}, \citenamefont {Heinz}, \citenamefont {Kim},
  \citenamefont {Pinczuk}, \citenamefont {Flynn},\ and\ \citenamefont
  {Pasupathy}}]{Zhao999}%
  \BibitemOpen
  \bibfield  {author} {\bibinfo {author} {\bibfnamefont {L.}~\bibnamefont
  {Zhao}}, \bibinfo {author} {\bibfnamefont {R.}~\bibnamefont {He}}, \bibinfo
  {author} {\bibfnamefont {K.~T.}\ \bibnamefont {Rim}}, \bibinfo {author}
  {\bibfnamefont {T.}~\bibnamefont {Schiros}}, \bibinfo {author} {\bibfnamefont
  {K.~S.}\ \bibnamefont {Kim}}, \bibinfo {author} {\bibfnamefont
  {H.}~\bibnamefont {Zhou}}, \bibinfo {author} {\bibfnamefont {C.}~\bibnamefont
  {Guti{\'e}rrez}}, \bibinfo {author} {\bibfnamefont {S.~P.}\ \bibnamefont
  {Chockalingam}}, \bibinfo {author} {\bibfnamefont {C.~J.}\ \bibnamefont
  {Arguello}}, \bibinfo {author} {\bibfnamefont {L.}~\bibnamefont
  {P{\'a}lov{\'a}}}, \bibinfo {author} {\bibfnamefont {D.}~\bibnamefont
  {Nordlund}}, \bibinfo {author} {\bibfnamefont {M.~S.}\ \bibnamefont
  {Hybertsen}}, \bibinfo {author} {\bibfnamefont {D.~R.}\ \bibnamefont
  {Reichman}}, \bibinfo {author} {\bibfnamefont {T.~F.}\ \bibnamefont {Heinz}},
  \bibinfo {author} {\bibfnamefont {P.}~\bibnamefont {Kim}}, \bibinfo {author}
  {\bibfnamefont {A.}~\bibnamefont {Pinczuk}}, \bibinfo {author} {\bibfnamefont
  {G.~W.}\ \bibnamefont {Flynn}}, \ and\ \bibinfo {author} {\bibfnamefont
  {A.~N.}\ \bibnamefont {Pasupathy}},\ }\href {\doibase
  10.1126/science.1208759} {\bibfield  {journal} {\bibinfo  {journal}
  {Science}\ }\textbf {\bibinfo {volume} {333}},\ \bibinfo {pages} {999}
  (\bibinfo {year} {2011})}\BibitemShut {NoStop}%
\bibitem [{\citenamefont {Gonz{\'a}lez-Herrero}\ \emph
  {et~al.}(2016)\citenamefont {Gonz{\'a}lez-Herrero}, \citenamefont
  {G{\'o}mez-Rodr{\'\i}guez}, \citenamefont {Mallet}, \citenamefont {Moaied},
  \citenamefont {Palacios}, \citenamefont {Salgado}, \citenamefont {Ugeda},
  \citenamefont {Veuillen}, \citenamefont {Yndurain},\ and\ \citenamefont
  {Brihuega}}]{Gonzalez-Herrero437}%
  \BibitemOpen
  \bibfield  {author} {\bibinfo {author} {\bibfnamefont {H.}~\bibnamefont
  {Gonz{\'a}lez-Herrero}}, \bibinfo {author} {\bibfnamefont {J.~M.}\
  \bibnamefont {G{\'o}mez-Rodr{\'\i}guez}}, \bibinfo {author} {\bibfnamefont
  {P.}~\bibnamefont {Mallet}}, \bibinfo {author} {\bibfnamefont
  {M.}~\bibnamefont {Moaied}}, \bibinfo {author} {\bibfnamefont {J.~J.}\
  \bibnamefont {Palacios}}, \bibinfo {author} {\bibfnamefont {C.}~\bibnamefont
  {Salgado}}, \bibinfo {author} {\bibfnamefont {M.~M.}\ \bibnamefont {Ugeda}},
  \bibinfo {author} {\bibfnamefont {J.-Y.}\ \bibnamefont {Veuillen}}, \bibinfo
  {author} {\bibfnamefont {F.}~\bibnamefont {Yndurain}}, \ and\ \bibinfo
  {author} {\bibfnamefont {I.}~\bibnamefont {Brihuega}},\ }\href {\doibase
  10.1126/science.aad8038} {\bibfield  {journal} {\bibinfo  {journal}
  {Science}\ }\textbf {\bibinfo {volume} {352}},\ \bibinfo {pages} {437}
  (\bibinfo {year} {2016})}\BibitemShut {NoStop}%
\bibitem [{\citenamefont {Wang}\ \emph {et~al.}(2012)\citenamefont {Wang},
  \citenamefont {Brar}, \citenamefont {Shytov}, \citenamefont {Wu},
  \citenamefont {Regan}, \citenamefont {Tsai}, \citenamefont {Zettl},
  \citenamefont {Levitov},\ and\ \citenamefont {Crommie}}]{Wang2012}%
  \BibitemOpen
  \bibfield  {author} {\bibinfo {author} {\bibfnamefont {Y.}~\bibnamefont
  {Wang}}, \bibinfo {author} {\bibfnamefont {V.~W.}\ \bibnamefont {Brar}},
  \bibinfo {author} {\bibfnamefont {A.~V.}\ \bibnamefont {Shytov}}, \bibinfo
  {author} {\bibfnamefont {Q.}~\bibnamefont {Wu}}, \bibinfo {author}
  {\bibfnamefont {W.}~\bibnamefont {Regan}}, \bibinfo {author} {\bibfnamefont
  {H.-Z.}\ \bibnamefont {Tsai}}, \bibinfo {author} {\bibfnamefont
  {A.}~\bibnamefont {Zettl}}, \bibinfo {author} {\bibfnamefont {L.~S.}\
  \bibnamefont {Levitov}}, \ and\ \bibinfo {author} {\bibfnamefont {M.~F.}\
  \bibnamefont {Crommie}},\ }\href {\doibase 10.1038/nphys2379} {\bibfield
  {journal} {\bibinfo  {journal} {Nature Phys.}\ }\textbf {\bibinfo {volume}
  {8}},\ \bibinfo {pages} {653} (\bibinfo {year} {2012})}\BibitemShut {NoStop}%
\bibitem [{\citenamefont {Brar}\ \emph {et~al.}(2011)\citenamefont {Brar},
  \citenamefont {Decker}, \citenamefont {Solowan}, \citenamefont {Wang},
  \citenamefont {Maserati}, \citenamefont {Chan}, \citenamefont {Lee},
  \citenamefont {Girit}, \citenamefont {Zettl}, \citenamefont {Louie},
  \citenamefont {Cohen},\ and\ \citenamefont {Crommie}}]{Brar2011}%
  \BibitemOpen
  \bibfield  {author} {\bibinfo {author} {\bibfnamefont {V.~W.}\ \bibnamefont
  {Brar}}, \bibinfo {author} {\bibfnamefont {R.}~\bibnamefont {Decker}},
  \bibinfo {author} {\bibfnamefont {H.-M.}\ \bibnamefont {Solowan}}, \bibinfo
  {author} {\bibfnamefont {Y.}~\bibnamefont {Wang}}, \bibinfo {author}
  {\bibfnamefont {L.}~\bibnamefont {Maserati}}, \bibinfo {author}
  {\bibfnamefont {K.~T.}\ \bibnamefont {Chan}}, \bibinfo {author}
  {\bibfnamefont {H.}~\bibnamefont {Lee}}, \bibinfo {author} {\bibfnamefont
  {C.~O.}\ \bibnamefont {Girit}}, \bibinfo {author} {\bibfnamefont
  {A.}~\bibnamefont {Zettl}}, \bibinfo {author} {\bibfnamefont {S.~G.}\
  \bibnamefont {Louie}}, \bibinfo {author} {\bibfnamefont {M.~L.}\ \bibnamefont
  {Cohen}}, \ and\ \bibinfo {author} {\bibfnamefont {M.~F.}\ \bibnamefont
  {Crommie}},\ }\href {\doibase 10.1038/nphys1807} {\bibfield  {journal}
  {\bibinfo  {journal} {Nature Phys.}\ }\textbf {\bibinfo {volume} {7}},\
  \bibinfo {pages} {43} (\bibinfo {year} {2011})}\BibitemShut {NoStop}%
\bibitem [{\citenamefont {Wang}\ \emph {et~al.}(2013)\citenamefont {Wang},
  \citenamefont {Wong}, \citenamefont {Shytov}, \citenamefont {Brar},
  \citenamefont {Choi}, \citenamefont {Wu}, \citenamefont {Tsai}, \citenamefont
  {Regan}, \citenamefont {Zettl}, \citenamefont {Kawakami}, \citenamefont
  {Louie}, \citenamefont {Levitov},\ and\ \citenamefont {Crommie}}]{Wang2013}%
  \BibitemOpen
  \bibfield  {author} {\bibinfo {author} {\bibfnamefont {Y.}~\bibnamefont
  {Wang}}, \bibinfo {author} {\bibfnamefont {D.}~\bibnamefont {Wong}}, \bibinfo
  {author} {\bibfnamefont {A.~V.}\ \bibnamefont {Shytov}}, \bibinfo {author}
  {\bibfnamefont {V.~W.}\ \bibnamefont {Brar}}, \bibinfo {author}
  {\bibfnamefont {S.}~\bibnamefont {Choi}}, \bibinfo {author} {\bibfnamefont
  {Q.}~\bibnamefont {Wu}}, \bibinfo {author} {\bibfnamefont {H.-Z.}\
  \bibnamefont {Tsai}}, \bibinfo {author} {\bibfnamefont {W.}~\bibnamefont
  {Regan}}, \bibinfo {author} {\bibfnamefont {A.}~\bibnamefont {Zettl}},
  \bibinfo {author} {\bibfnamefont {R.~K.}\ \bibnamefont {Kawakami}}, \bibinfo
  {author} {\bibfnamefont {S.~G.}\ \bibnamefont {Louie}}, \bibinfo {author}
  {\bibfnamefont {L.~S.}\ \bibnamefont {Levitov}}, \ and\ \bibinfo {author}
  {\bibfnamefont {M.~F.}\ \bibnamefont {Crommie}},\ }\href {\doibase
  10.1126/science.1234320} {\bibfield  {journal} {\bibinfo  {journal}
  {Science}\ }\textbf {\bibinfo {volume} {340}},\ \bibinfo {pages} {734}
  (\bibinfo {year} {2013})}\BibitemShut {NoStop}%
\bibitem [{\citenamefont {Luican-Mayer}\ \emph {et~al.}(2014)\citenamefont
  {Luican-Mayer}, \citenamefont {Kharitonov}, \citenamefont {Li}, \citenamefont
  {Lu}, \citenamefont {Skachko}, \citenamefont {Gon\ifmmode~\mbox{\c{c}}\else
  \c{c}\fi{}alves}, \citenamefont {Watanabe}, \citenamefont {Taniguchi},\ and\
  \citenamefont {Andrei}}]{PhysRevLett.112.036804}%
  \BibitemOpen
  \bibfield  {author} {\bibinfo {author} {\bibfnamefont {A.}~\bibnamefont
  {Luican-Mayer}}, \bibinfo {author} {\bibfnamefont {M.}~\bibnamefont
  {Kharitonov}}, \bibinfo {author} {\bibfnamefont {G.}~\bibnamefont {Li}},
  \bibinfo {author} {\bibfnamefont {C.-P.}\ \bibnamefont {Lu}}, \bibinfo
  {author} {\bibfnamefont {I.}~\bibnamefont {Skachko}}, \bibinfo {author}
  {\bibfnamefont {A.-M.~B.}\ \bibnamefont {Gon\ifmmode~\mbox{\c{c}}\else
  \c{c}\fi{}alves}}, \bibinfo {author} {\bibfnamefont {K.}~\bibnamefont
  {Watanabe}}, \bibinfo {author} {\bibfnamefont {T.}~\bibnamefont {Taniguchi}},
  \ and\ \bibinfo {author} {\bibfnamefont {E.~Y.}\ \bibnamefont {Andrei}},\
  }\href {\doibase 10.1103/PhysRevLett.112.036804} {\bibfield  {journal}
  {\bibinfo  {journal} {Phys. Rev. Lett.}\ }\textbf {\bibinfo {volume} {112}},\
  \bibinfo {pages} {036804} (\bibinfo {year} {2014})}\BibitemShut {NoStop}%
\bibitem [{\citenamefont {Mao}\ \emph {et~al.}(2016)\citenamefont {Mao},
  \citenamefont {Jiang}, \citenamefont {Moldovan}, \citenamefont {Li},
  \citenamefont {Watanabe}, \citenamefont {Taniguchi}, \citenamefont {Masir},
  \citenamefont {Peeters},\ and\ \citenamefont {Andrei}}]{Mao2016}%
  \BibitemOpen
  \bibfield  {author} {\bibinfo {author} {\bibfnamefont {J.}~\bibnamefont
  {Mao}}, \bibinfo {author} {\bibfnamefont {Y.}~\bibnamefont {Jiang}}, \bibinfo
  {author} {\bibfnamefont {D.}~\bibnamefont {Moldovan}}, \bibinfo {author}
  {\bibfnamefont {G.}~\bibnamefont {Li}}, \bibinfo {author} {\bibfnamefont
  {K.}~\bibnamefont {Watanabe}}, \bibinfo {author} {\bibfnamefont
  {T.}~\bibnamefont {Taniguchi}}, \bibinfo {author} {\bibfnamefont {M.~R.}\
  \bibnamefont {Masir}}, \bibinfo {author} {\bibfnamefont {F.~M.}\ \bibnamefont
  {Peeters}}, \ and\ \bibinfo {author} {\bibfnamefont {E.~Y.}\ \bibnamefont
  {Andrei}},\ }\href {\doibase 10.1038/nphys3665} {\bibfield  {journal}
  {\bibinfo  {journal} {Nature Phys.}\ }\textbf {\bibinfo {volume} {12}},\
  \bibinfo {pages} {545} (\bibinfo {year} {2016})}\BibitemShut {NoStop}%
\bibitem [{\citenamefont {Chan}\ \emph {et~al.}(2008)\citenamefont {Chan},
  \citenamefont {Neaton},\ and\ \citenamefont {Cohen}}]{Chan2008}%
  \BibitemOpen
  \bibfield  {author} {\bibinfo {author} {\bibfnamefont {K.~T.}\ \bibnamefont
  {Chan}}, \bibinfo {author} {\bibfnamefont {J.~B.}\ \bibnamefont {Neaton}}, \
  and\ \bibinfo {author} {\bibfnamefont {M.~L.}\ \bibnamefont {Cohen}},\ }\href
  {\doibase 10.1103/PhysRevB.77.235430} {\bibfield  {journal} {\bibinfo
  {journal} {Phys. Rev. B}\ }\textbf {\bibinfo {volume} {77}},\ \bibinfo
  {pages} {235430} (\bibinfo {year} {2008})}\BibitemShut {NoStop}%
\bibitem [{\citenamefont {Parq}\ \emph {et~al.}(2010)\citenamefont {Parq},
  \citenamefont {Yu}, \citenamefont {Kwon},\ and\ \citenamefont
  {Kim}}]{Parq2010}%
  \BibitemOpen
  \bibfield  {author} {\bibinfo {author} {\bibfnamefont {J.-H.}\ \bibnamefont
  {Parq}}, \bibinfo {author} {\bibfnamefont {J.}~\bibnamefont {Yu}}, \bibinfo
  {author} {\bibfnamefont {Y.-K.}\ \bibnamefont {Kwon}}, \ and\ \bibinfo
  {author} {\bibfnamefont {G.}~\bibnamefont {Kim}},\ }\href {\doibase
  10.1103/PhysRevB.82.193406} {\bibfield  {journal} {\bibinfo  {journal} {Phys.
  Rev. B}\ }\textbf {\bibinfo {volume} {82}},\ \bibinfo {pages} {193406}
  (\bibinfo {year} {2010})}\BibitemShut {NoStop}%
\bibitem [{\citenamefont {Chan}\ \emph {et~al.}(2011)\citenamefont {Chan},
  \citenamefont {Lee},\ and\ \citenamefont {Cohen}}]{Chan2011}%
  \BibitemOpen
  \bibfield  {author} {\bibinfo {author} {\bibfnamefont {K.~T.}\ \bibnamefont
  {Chan}}, \bibinfo {author} {\bibfnamefont {H.}~\bibnamefont {Lee}}, \ and\
  \bibinfo {author} {\bibfnamefont {M.~L.}\ \bibnamefont {Cohen}},\ }\href
  {\doibase 10.1103/PhysRevB.84.165419} {\bibfield  {journal} {\bibinfo
  {journal} {Phys. Rev. B}\ }\textbf {\bibinfo {volume} {84}},\ \bibinfo
  {pages} {165419} (\bibinfo {year} {2011})}\BibitemShut {NoStop}%
\bibitem [{\citenamefont {Sun}\ \emph {et~al.}(2015)\citenamefont {Sun},
  \citenamefont {Tang}, \citenamefont {Ren}, \citenamefont {Wang},
  \citenamefont {JinYu}, \citenamefont {Du},\ and\ \citenamefont
  {Zhang}}]{Sun2015668}%
  \BibitemOpen
  \bibfield  {author} {\bibinfo {author} {\bibfnamefont {M.}~\bibnamefont
  {Sun}}, \bibinfo {author} {\bibfnamefont {W.}~\bibnamefont {Tang}}, \bibinfo
  {author} {\bibfnamefont {Q.}~\bibnamefont {Ren}}, \bibinfo {author}
  {\bibfnamefont {S.}~\bibnamefont {Wang}}, \bibinfo {author} {\bibnamefont
  {JinYu}}, \bibinfo {author} {\bibfnamefont {Y.}~\bibnamefont {Du}}, \ and\
  \bibinfo {author} {\bibfnamefont {Y.}~\bibnamefont {Zhang}},\ }\href
  {\doibase 10.1016/j.apsusc.2015.08.102} {\bibfield  {journal} {\bibinfo
  {journal} {Appl. Surf. Sci.}\ }\textbf {\bibinfo {volume} {356}},\ \bibinfo
  {pages} {668} (\bibinfo {year} {2015})}\BibitemShut {NoStop}%
\bibitem [{\citenamefont {Pereira}\ \emph {et~al.}(2007)\citenamefont
  {Pereira}, \citenamefont {Nilsson},\ and\ \citenamefont
  {Castro~Neto}}]{Pereira2007}%
  \BibitemOpen
  \bibfield  {author} {\bibinfo {author} {\bibfnamefont {V.~M.}\ \bibnamefont
  {Pereira}}, \bibinfo {author} {\bibfnamefont {J.}~\bibnamefont {Nilsson}}, \
  and\ \bibinfo {author} {\bibfnamefont {A.~H.}\ \bibnamefont {Castro~Neto}},\
  }\href {\doibase 10.1103/PhysRevLett.99.166802} {\bibfield  {journal}
  {\bibinfo  {journal} {Phys. Rev. Lett.}\ }\textbf {\bibinfo {volume} {99}},\
  \bibinfo {pages} {166802} (\bibinfo {year} {2007})}\BibitemShut {NoStop}%
\bibitem [{\citenamefont {Polini}\ \emph {et~al.}(2008)\citenamefont {Polini},
  \citenamefont {Tomadin}, \citenamefont {Asgari},\ and\ \citenamefont
  {MacDonald}}]{Polini2008}%
  \BibitemOpen
  \bibfield  {author} {\bibinfo {author} {\bibfnamefont {M.}~\bibnamefont
  {Polini}}, \bibinfo {author} {\bibfnamefont {A.}~\bibnamefont {Tomadin}},
  \bibinfo {author} {\bibfnamefont {R.}~\bibnamefont {Asgari}}, \ and\ \bibinfo
  {author} {\bibfnamefont {A.~H.}\ \bibnamefont {MacDonald}},\ }\href {\doibase
  10.1103/PhysRevB.78.115426} {\bibfield  {journal} {\bibinfo  {journal} {Phys.
  Rev. B}\ }\textbf {\bibinfo {volume} {78}},\ \bibinfo {pages} {115426}
  (\bibinfo {year} {2008})}\BibitemShut {NoStop}%
\bibitem [{\citenamefont {Lu}\ \emph {et~al.}(2015)\citenamefont {Lu},
  \citenamefont {Moroz},\ and\ \citenamefont {Muratov}}]{Lu2015}%
  \BibitemOpen
  \bibfield  {author} {\bibinfo {author} {\bibfnamefont {J.}~\bibnamefont
  {Lu}}, \bibinfo {author} {\bibfnamefont {V.}~\bibnamefont {Moroz}}, \ and\
  \bibinfo {author} {\bibfnamefont {C.~B.}\ \bibnamefont {Muratov}},\ }\href
  {\doibase 10.1007/s00332-015-9259-4} {\bibfield  {journal} {\bibinfo
  {journal} {J. Nonlinear Sci.}\ }\textbf {\bibinfo {volume} {25}},\ \bibinfo
  {pages} {1391} (\bibinfo {year} {2015})}\BibitemShut {NoStop}%
\bibitem [{\citenamefont {Novikov}(2007)}]{Novikov2007}%
  \BibitemOpen
  \bibfield  {author} {\bibinfo {author} {\bibfnamefont {D.~S.}\ \bibnamefont
  {Novikov}},\ }\href {\doibase 10.1103/PhysRevB.76.245435} {\bibfield
  {journal} {\bibinfo  {journal} {Phys. Rev. B}\ }\textbf {\bibinfo {volume}
  {76}},\ \bibinfo {pages} {245435} (\bibinfo {year} {2007})}\BibitemShut
  {NoStop}%
\bibitem [{\citenamefont {Ando}(2006)}]{Ando2006}%
  \BibitemOpen
  \bibfield  {author} {\bibinfo {author} {\bibfnamefont {T.}~\bibnamefont
  {Ando}},\ }\href {\doibase 10.1143/JPSJ.75.074716} {\bibfield  {journal}
  {\bibinfo  {journal} {J. Phys. Soc. Japan}\ }\textbf {\bibinfo {volume}
  {75}},\ \bibinfo {pages} {074716} (\bibinfo {year} {2006})}\BibitemShut
  {NoStop}%
\bibitem [{\citenamefont {Hwang}\ \emph {et~al.}(2007)\citenamefont {Hwang},
  \citenamefont {Adam},\ and\ \citenamefont {Das~Sarma}}]{Hwang2007}%
  \BibitemOpen
  \bibfield  {author} {\bibinfo {author} {\bibfnamefont {E.~H.}\ \bibnamefont
  {Hwang}}, \bibinfo {author} {\bibfnamefont {S.}~\bibnamefont {Adam}}, \ and\
  \bibinfo {author} {\bibfnamefont {S.}~\bibnamefont {Das~Sarma}},\ }\href
  {\doibase 10.1103/PhysRevLett.98.186806} {\bibfield  {journal} {\bibinfo
  {journal} {Phys. Rev. Lett.}\ }\textbf {\bibinfo {volume} {98}},\ \bibinfo
  {pages} {186806} (\bibinfo {year} {2007})}\BibitemShut {NoStop}%
\bibitem [{\citenamefont {Katsnelson}(2006)}]{Katsnelson2006}%
  \BibitemOpen
  \bibfield  {author} {\bibinfo {author} {\bibfnamefont {M.~I.}\ \bibnamefont
  {Katsnelson}},\ }\href {\doibase 10.1103/PhysRevB.74.201401} {\bibfield
  {journal} {\bibinfo  {journal} {Phys. Rev. B}\ }\textbf {\bibinfo {volume}
  {74}},\ \bibinfo {pages} {201401} (\bibinfo {year} {2006})}\BibitemShut
  {NoStop}%
\bibitem [{\citenamefont {Fogler}\ \emph {et~al.}(2007)\citenamefont {Fogler},
  \citenamefont {Novikov},\ and\ \citenamefont {Shklovskii}}]{Fogler2007}%
  \BibitemOpen
  \bibfield  {author} {\bibinfo {author} {\bibfnamefont {M.~M.}\ \bibnamefont
  {Fogler}}, \bibinfo {author} {\bibfnamefont {D.~S.}\ \bibnamefont {Novikov}},
  \ and\ \bibinfo {author} {\bibfnamefont {B.~I.}\ \bibnamefont {Shklovskii}},\
  }\href {\doibase 10.1103/PhysRevB.76.233402} {\bibfield  {journal} {\bibinfo
  {journal} {Phys. Rev. B}\ }\textbf {\bibinfo {volume} {76}},\ \bibinfo
  {pages} {233402} (\bibinfo {year} {2007})}\BibitemShut {NoStop}%
\bibitem [{\citenamefont {Terekhov}\ \emph {et~al.}(2008)\citenamefont
  {Terekhov}, \citenamefont {Milstein}, \citenamefont {Kotov},\ and\
  \citenamefont {Sushkov}}]{Terekhov2008}%
  \BibitemOpen
  \bibfield  {author} {\bibinfo {author} {\bibfnamefont {I.~S.}\ \bibnamefont
  {Terekhov}}, \bibinfo {author} {\bibfnamefont {A.~I.}\ \bibnamefont
  {Milstein}}, \bibinfo {author} {\bibfnamefont {V.~N.}\ \bibnamefont {Kotov}},
  \ and\ \bibinfo {author} {\bibfnamefont {O.~P.}\ \bibnamefont {Sushkov}},\
  }\href {\doibase 10.1103/PhysRevLett.100.076803} {\bibfield  {journal}
  {\bibinfo  {journal} {Phys. Rev. Lett.}\ }\textbf {\bibinfo {volume} {100}},\
  \bibinfo {pages} {076803} (\bibinfo {year} {2008})}\BibitemShut {NoStop}%
\bibitem [{\citenamefont {Wong}\ \emph {et~al.}()\citenamefont {Wong},
  \citenamefont {Corsetti}, \citenamefont {Wang}, \citenamefont {Brar},
  \citenamefont {Tsai}, \citenamefont {Wu}, \citenamefont {Kawakami},
  \citenamefont {Zettl}, \citenamefont {Mostofi}, \citenamefont {Lischner},\
  and\ \citenamefont {Crommie}}]{WongCa}%
  \BibitemOpen
  \bibfield  {author} {\bibinfo {author} {\bibfnamefont {D.}~\bibnamefont
  {Wong}}, \bibinfo {author} {\bibfnamefont {F.}~\bibnamefont {Corsetti}},
  \bibinfo {author} {\bibfnamefont {Y.}~\bibnamefont {Wang}}, \bibinfo {author}
  {\bibfnamefont {V.~W.}\ \bibnamefont {Brar}}, \bibinfo {author}
  {\bibfnamefont {H.-Z.}\ \bibnamefont {Tsai}}, \bibinfo {author}
  {\bibfnamefont {Q.}~\bibnamefont {Wu}}, \bibinfo {author} {\bibfnamefont
  {R.~K.}\ \bibnamefont {Kawakami}}, \bibinfo {author} {\bibfnamefont
  {A.}~\bibnamefont {Zettl}}, \bibinfo {author} {\bibfnamefont {A.~A.}\
  \bibnamefont {Mostofi}}, \bibinfo {author} {\bibfnamefont {J.}~\bibnamefont
  {Lischner}}, \ and\ \bibinfo {author} {\bibfnamefont {M.~F.}\ \bibnamefont
  {Crommie}},\ }\href@noop {} {\enquote {\bibinfo {title} {Spatially resolving
  density-dependent screening around a single charged atom in graphene},}\
  }\bibinfo {note} {Submitted}\BibitemShut {NoStop}%
\bibitem [{\citenamefont {Bell}\ \emph {et~al.}(2014)\citenamefont {Bell},
  \citenamefont {Payne},\ and\ \citenamefont {Mostofi}}]{Bell2014}%
  \BibitemOpen
  \bibfield  {author} {\bibinfo {author} {\bibfnamefont {R.~A.}\ \bibnamefont
  {Bell}}, \bibinfo {author} {\bibfnamefont {M.~C.}\ \bibnamefont {Payne}}, \
  and\ \bibinfo {author} {\bibfnamefont {A.~A.}\ \bibnamefont {Mostofi}},\
  }\href {\doibase 10.1063/1.4898712} {\bibfield  {journal} {\bibinfo
  {journal} {J. Chem. Phys.}\ }\textbf {\bibinfo {volume} {141}},\ \bibinfo
  {pages} {164703} (\bibinfo {year} {2014})}\BibitemShut {NoStop}%
\bibitem [{\citenamefont {Shung}(1986)}]{Shung1986}%
  \BibitemOpen
  \bibfield  {author} {\bibinfo {author} {\bibfnamefont {K.~W.-K.}\
  \bibnamefont {Shung}},\ }\href {\doibase 10.1103/PhysRevB.34.979} {\bibfield
  {journal} {\bibinfo  {journal} {Phys. Rev. B}\ }\textbf {\bibinfo {volume}
  {34}},\ \bibinfo {pages} {979} (\bibinfo {year} {1986})}\BibitemShut
  {NoStop}%
\bibitem [{\citenamefont {Wunsch}\ \emph {et~al.}(2006)\citenamefont {Wunsch},
  \citenamefont {Stauber}, \citenamefont {Sols},\ and\ \citenamefont
  {Guinea}}]{Wunsch2006}%
  \BibitemOpen
  \bibfield  {author} {\bibinfo {author} {\bibfnamefont {B.}~\bibnamefont
  {Wunsch}}, \bibinfo {author} {\bibfnamefont {T.}~\bibnamefont {Stauber}},
  \bibinfo {author} {\bibfnamefont {F.}~\bibnamefont {Sols}}, \ and\ \bibinfo
  {author} {\bibfnamefont {F.}~\bibnamefont {Guinea}},\ }\href {\doibase
  10.1088/1367-2630/8/12/318} {\bibfield  {journal} {\bibinfo  {journal} {New
  J. Phys.}\ }\textbf {\bibinfo {volume} {8}},\ \bibinfo {pages} {318}
  (\bibinfo {year} {2006})}\BibitemShut {NoStop}%
\bibitem [{\citenamefont {Hwang}\ and\ \citenamefont
  {Das~Sarma}(2007)}]{Hwang2007b}%
  \BibitemOpen
  \bibfield  {author} {\bibinfo {author} {\bibfnamefont {E.~H.}\ \bibnamefont
  {Hwang}}\ and\ \bibinfo {author} {\bibfnamefont {S.}~\bibnamefont
  {Das~Sarma}},\ }\href {\doibase 10.1103/PhysRevB.75.205418} {\bibfield
  {journal} {\bibinfo  {journal} {Phys. Rev. B}\ }\textbf {\bibinfo {volume}
  {75}},\ \bibinfo {pages} {205418} (\bibinfo {year} {2007})}\BibitemShut
  {NoStop}%
\bibitem [{\citenamefont {Wehling}\ \emph {et~al.}(2011)\citenamefont
  {Wehling}, \citenamefont {\ifmmode \mbox{\c{S}}\else \c{S}\fi{}a\ifmmode
  \mbox{\c{s}}\else \c{s}\fi{}\ifmmode \imath \else \i
  \fi{}o\ifmmode~\breve{g}\else \u{g}\fi{}lu}, \citenamefont {Friedrich},
  \citenamefont {Lichtenstein}, \citenamefont {Katsnelson},\ and\ \citenamefont
  {Bl\"ugel}}]{Wehling2011}%
  \BibitemOpen
  \bibfield  {author} {\bibinfo {author} {\bibfnamefont {T.~O.}\ \bibnamefont
  {Wehling}}, \bibinfo {author} {\bibfnamefont {E.}~\bibnamefont {\ifmmode
  \mbox{\c{S}}\else \c{S}\fi{}a\ifmmode \mbox{\c{s}}\else \c{s}\fi{}\ifmmode
  \imath \else \i \fi{}o\ifmmode~\breve{g}\else \u{g}\fi{}lu}}, \bibinfo
  {author} {\bibfnamefont {C.}~\bibnamefont {Friedrich}}, \bibinfo {author}
  {\bibfnamefont {A.~I.}\ \bibnamefont {Lichtenstein}}, \bibinfo {author}
  {\bibfnamefont {M.~I.}\ \bibnamefont {Katsnelson}}, \ and\ \bibinfo {author}
  {\bibfnamefont {S.}~\bibnamefont {Bl\"ugel}},\ }\href {\doibase
  10.1103/PhysRevLett.106.236805} {\bibfield  {journal} {\bibinfo  {journal}
  {Phys. Rev. Lett.}\ }\textbf {\bibinfo {volume} {106}},\ \bibinfo {pages}
  {236805} (\bibinfo {year} {2011})}\BibitemShut {NoStop}%
\bibitem [{\citenamefont {Perdew}\ \emph {et~al.}(1996)\citenamefont {Perdew},
  \citenamefont {Burke},\ and\ \citenamefont {Ernzerhof}}]{PBE}%
  \BibitemOpen
  \bibfield  {author} {\bibinfo {author} {\bibfnamefont {J.~P.}\ \bibnamefont
  {Perdew}}, \bibinfo {author} {\bibfnamefont {K.}~\bibnamefont {Burke}}, \
  and\ \bibinfo {author} {\bibfnamefont {M.}~\bibnamefont {Ernzerhof}},\ }\href
  {\doibase 10.1103/PhysRevLett.77.3865} {\bibfield  {journal} {\bibinfo
  {journal} {Phys. Rev. Lett.}\ }\textbf {\bibinfo {volume} {77}},\ \bibinfo
  {pages} {3865} (\bibinfo {year} {1996})}\BibitemShut {NoStop}%
\bibitem [{\citenamefont {Clark}\ \emph {et~al.}(2005)\citenamefont {Clark},
  \citenamefont {Segall}, \citenamefont {Pickard}, \citenamefont {Hasnip},
  \citenamefont {Probert}, \citenamefont {Refson},\ and\ \citenamefont
  {Payne}}]{clark2005first}%
  \BibitemOpen
  \bibfield  {author} {\bibinfo {author} {\bibfnamefont {S.~J.}\ \bibnamefont
  {Clark}}, \bibinfo {author} {\bibfnamefont {M.~D.}\ \bibnamefont {Segall}},
  \bibinfo {author} {\bibfnamefont {C.~J.}\ \bibnamefont {Pickard}}, \bibinfo
  {author} {\bibfnamefont {P.~J.}\ \bibnamefont {Hasnip}}, \bibinfo {author}
  {\bibfnamefont {M.~I.~J.}\ \bibnamefont {Probert}}, \bibinfo {author}
  {\bibfnamefont {K.}~\bibnamefont {Refson}}, \ and\ \bibinfo {author}
  {\bibfnamefont {M.~C.}\ \bibnamefont {Payne}},\ }\href {\doibase
  10.1524/zkri.220.5.567.65075} {\bibfield  {journal} {\bibinfo  {journal} {Z.
  Kristallogr. Cryst. Mater.}\ }\textbf {\bibinfo {volume} {220}},\ \bibinfo
  {pages} {567} (\bibinfo {year} {2005})}\BibitemShut {NoStop}%
\bibitem [{\citenamefont {Skylaris}\ \emph {et~al.}(2005)\citenamefont
  {Skylaris}, \citenamefont {Haynes}, \citenamefont {Mostofi},\ and\
  \citenamefont {Payne}}]{ONETEP}%
  \BibitemOpen
  \bibfield  {author} {\bibinfo {author} {\bibfnamefont {C.-K.}\ \bibnamefont
  {Skylaris}}, \bibinfo {author} {\bibfnamefont {P.~D.}\ \bibnamefont
  {Haynes}}, \bibinfo {author} {\bibfnamefont {A.~A.}\ \bibnamefont {Mostofi}},
  \ and\ \bibinfo {author} {\bibfnamefont {M.~C.}\ \bibnamefont {Payne}},\
  }\href {\doibase 10.1063/1.1839852} {\bibfield  {journal} {\bibinfo
  {journal} {J. Chem. Phys.}\ }\textbf {\bibinfo {volume} {122}},\ \bibinfo
  {pages} {084119} (\bibinfo {year} {2005})}\BibitemShut {NoStop}%
\bibitem [{\citenamefont {Hine}\ \emph {et~al.}(2012)\citenamefont {Hine},
  \citenamefont {Avraam}, \citenamefont {Tangney},\ and\ \citenamefont
  {Haynes}}]{LDOS}%
  \BibitemOpen
  \bibfield  {author} {\bibinfo {author} {\bibfnamefont {N.~D.~M.}\
  \bibnamefont {Hine}}, \bibinfo {author} {\bibfnamefont {P.~W.}\ \bibnamefont
  {Avraam}}, \bibinfo {author} {\bibfnamefont {P.}~\bibnamefont {Tangney}}, \
  and\ \bibinfo {author} {\bibfnamefont {P.~D.}\ \bibnamefont {Haynes}},\
  }\href {\doibase 10.1088/1742-6596/367/1/012002} {\bibfield  {journal}
  {\bibinfo  {journal} {J. Phys.: Conf. Ser.}\ }\textbf {\bibinfo {volume}
  {367}},\ \bibinfo {pages} {012002} (\bibinfo {year} {2012})}\BibitemShut
  {NoStop}%
\bibitem [{\citenamefont {Ruiz-Serrano}\ and\ \citenamefont
  {Skylaris}(2013)}]{EDFT}%
  \BibitemOpen
  \bibfield  {author} {\bibinfo {author} {\bibfnamefont {A.}~\bibnamefont
  {Ruiz-Serrano}}\ and\ \bibinfo {author} {\bibfnamefont {C.-K.}\ \bibnamefont
  {Skylaris}},\ }\href {\doibase 10.1063/1.4817001} {\bibfield  {journal}
  {\bibinfo  {journal} {J. Chem. Phys.}\ }\textbf {\bibinfo {volume} {139}},\
  \bibinfo {pages} {054107} (\bibinfo {year} {2013})}\BibitemShut {NoStop}%
\bibitem [{gra()}]{graphene-code}%
  \BibitemOpen
  \href@noop {} {}\bibinfo {note}
  {\url{http://www.cmth.ph.ic.ac.uk/graphene/}}\BibitemShut {NoStop}%
\bibitem [{\citenamefont {DiVincenzo}\ and\ \citenamefont
  {Mele}(1984)}]{DiVincenzo1984}%
  \BibitemOpen
  \bibfield  {author} {\bibinfo {author} {\bibfnamefont {D.~P.}\ \bibnamefont
  {DiVincenzo}}\ and\ \bibinfo {author} {\bibfnamefont {E.~J.}\ \bibnamefont
  {Mele}},\ }\href {\doibase 10.1103/PhysRevB.29.1685} {\bibfield  {journal}
  {\bibinfo  {journal} {Phys. Rev. B}\ }\textbf {\bibinfo {volume} {29}},\
  \bibinfo {pages} {1685} (\bibinfo {year} {1984})}\BibitemShut {NoStop}%
\bibitem [{Note1()}]{Note1}%
  \BibitemOpen
  \bibinfo {note} {However, the four models result in somewhat lower overall
  values of $\protect \mathcal {F}$ than fits based on empirical forms for the
  screened potential, discussed in Fig.~S3 in the Supplementary
  Material.}\BibitemShut {Stop}%
\bibitem [{fig()}]{figshare-files}%
  \BibitemOpen
  \href@noop {} {}\bibinfo {note}
  {\url{https://dx.doi.org/10.6084/m9.figshare.3824451}}\BibitemShut {NoStop}%
\end{thebibliography}
\end{document}